\documentclass[aps,prl,showpacs,twocolumn,preprintnumbers,superscriptaddress,nofootinbib,floatfix,10pt]{revtex4-1}
\usepackage{amsfonts,amssymb,stmaryrd,latexsym,amsmath,braket}
\usepackage{graphicx,subfigure}
\usepackage{comment}
\usepackage{times}
\usepackage{slashed}
\usepackage{bm}
\usepackage{braket}
\usepackage{mathrsfs}

\makeatletter
\let\saved@includegraphics\includegraphics
\AtBeginDocument{\let\includegraphics\saved@includegraphics}
\renewenvironment*{figure}{\@float{figure}}{\end@float}
\makeatother

\begin{document}

\bibliographystyle{naturemag}

\title{How machine learning conquers the unitary limit}

\author{Bastian Kaspschak}

\affiliation{Helmholtz-Institut f\"ur Strahlen- und Kernphysik and Bethe
Center
for Theoretical Physics, Universit\"at Bonn, D-53115 Bonn, Germany}

\author{Ulf-G. Mei{\ss}ner}

\affiliation{Helmholtz-Institut f\"ur Strahlen- und Kernphysik and Bethe
Center
for Theoretical Physics, Universit\"at Bonn, D-53115 Bonn, Germany}

\affiliation{Institute for Advanced Simulation, Institut f\"ur Kernphysik,
and
J\"ulich Center for Hadron Physics, Forschungszentrum J\"ulich,
D-52425 J\"ulich, Germany}

\affiliation{Tbilisi State University, 0186 Tbilisi, Georgia}

\begin{abstract}
Machine learning has become a premier tool in physics and  other fields of science.
It has been shown that the quantum mechanical scattering problem can not only be solved with such
techniques, but  it was argued that the underlying neural network develops the Born series
for shallow potentials. However, classical machine learning algorithms fail in the unitary limit of an
infinite scattering length and vanishing effective range parameters. The unitary limit plays an
important role in our understanding of bound strongly interacting fermionic systems
and can be realized in cold atom experiments. Here, we develop a formalism that explains the unitary
limit in terms of what we define as unitary limit surfaces. This not only allows to investigate the unitary
limit geometrically in potential space, but also provides a numerically simple approach towards unnaturally
large scattering lengths with standard multilayer perceptrons. Its scope is therefore not limited to
applications in nuclear and atomic physics, but includes all systems that exhibit an unnaturally large scale.
\end{abstract}
\maketitle

\textit{Introduction:}~After neural networks have already been successfully used in experimental applications, such
as particle identification, see e.g.~\cite{Radovic:2018}, much progress has been made in recent years by applying
them to various fields of theoretical
physics, such as Refs.~\cite{Richards:2011za,Graff:2013cla,Buckley:2011kc,Carleo:2017,Mills:2017,Wetzel:2017ooo,He:2017set,Fujimoto:2017cdo,Wu:2018,Niu:2018trk,Brehmer:2018kdj,Steinheimer:2019iso,Larkoski:2017jix}. An interesting
property of neural networks is that their prediction is exclusively achieved in terms of simple mathematical
operations, especially matrix multiplications. Therefore, a neural network approach bypasses the underlying
mathematical framework of the respective theory and still provides satisfactory results. Despite their excellent
performance, a major drawback of many neural networks is their lack of interpretability, which is the reason why
neural networks are often referred to as ``black boxes''. However, there are methods to restore interpretability.
A premier example for this is given by~\cite{Wu:2018}: By investigating patterns in the networks' weights, it was
demonstrated that multilayer perceptrons (MLPs) develop perturbation theory in terms of the Born approximation in
order to predict natural S-wave scattering lengths $a_0$ for shallow potentials. Nevertheless this approach
fails for deeper potentials, especially if they give rise to zero-energy bound states and thereby to the unitary
limit $a_0\to \infty$. The physical reason for this is that the unitary limit is a highly non-perturbative scenario,
and in addition, the technical difficulty of reproducing a singularity by a neural network arises, which
requires unconventional architectures and training algorithms. We note that the unitary limit plays an
important role in our understanding of bound strongly interacting fermionic
systems~\cite{Efimov:1970zz,Heiselberg:2000bm,Braaten:2004rn,Bulgac:2005pj,Lee:2005fk,Konig:2016utl}
and can be realized in cold atom experiments, see, e.g.,~\cite{Grimm}. Therefore, the question arises how to deal with such
a scenario in terms of machine learning? Our idea is to analyze the unitary limit in potential space.
Therefore, we  develop a formalism that explains it as a movable singularity in potential space. This formalism
introduces two geometric quantities $f$ and $b_0$ that are regular in the unitary limit and therefore can be easily learned
by standard MLPs. Finally, unnatural as well as natural scattering lengths can be accurately predicted
by composing the respective networks.

\begin{figure}[t]
\includegraphics[width=0.9\columnwidth]{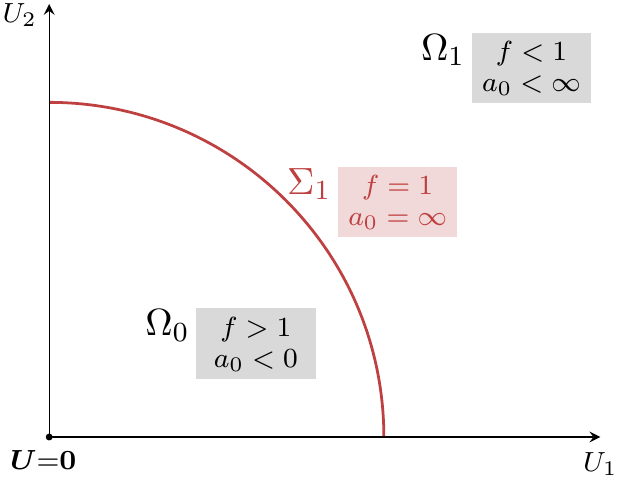}
\caption{Sketch of the regions $\Omega_0$ and $\Omega_1$ and the first unitary limit surface $\Sigma_1\subset\Omega_1$
for the degree $d=2$ of discretization. In this specific case, the potential space $\Omega$ is the first
quadrant of $\mathbb{R}^2$ and unitary limit surfaces are one-dimensional manifolds.}
\label{fig-0}
\vspace{-2mm}
\end{figure}

\textit{Discretized potentials and unitary limit surfaces:}~We aim at investigating the unitary limit, which is
why only attractive potentials need to be considered. For simplicity, the following analysis is restricted
to non-positive, spherically~symmetric~potentials $V(r)$~$\leq$~$0$ with finite range $\rho$.
Together with the~reduced~mass $\mu$, the latter parameterizes all dimensionless quantities. The most relevant ones
for describing low-energy S-wave scattering processes turn out to be the dimensionless potential
$U$~$=$~$-2\mu\rho^2~V$~$\geq$~$0$ and the S-wave scattering length~$a_0$. An important first step is to
discretize the potentials, since these can then be treated as vectors $\bm{U}$~$\in$~$\Omega$~$\subset$~$\mathbb{R}^d$
with non-negative components $U_n$~$=$~$U(n\rho/d)$~$\geq$~$0$ and become processable by common neural network
architectures, for details, see~\cite{SM}.
The degree $d$ of discretization thereby controls the granularity and corresponds to the inverse
step size of the emerging discretized potentials. As a further result of discretization, the domain of all
considered potentials is reduced to the first hyperoctant $\Omega$ of $\mathbb{R}^d$. Counting bound
states naturally splits the potential space $\Omega$~$=$~$\bigcup_{i\in\mathbb{N}_0}$~$\Omega_i$ into
pairwise disjunct, half-open regions~$\Omega_i$, with $\Omega_i$ containing all potentials that give
rise to exactly $i$ bound states. The $d$~$-$~$1$ dimensional hypersurface between two neighboring regions
\vspace{-0.2cm}
\begin{equation}
\Sigma_i \equiv \partial\Omega_{i-1}\cap \Omega_{i}~,
\vspace{-0.3cm}
\end{equation}
with $\Sigma_i$~$\subset$~$\Omega_{i}$ thereby consisting of all potentials $\bm{U}$~$\in$~$\Omega_i$ with a
zero-energy bound state, see Fig.~\ref{fig-0}. Since we observe the unitary~limit~$a_0$~$\to$~$\infty$ in this
scenario, we refer to $\Sigma_i$ as the $i^\text{th}$~unitary limit surface in $\Omega$. Considering the
scattering length as a function $a_0:\Omega\to\mathbb{R}$, this suggests a movable singularity on each unitary
limit surface. For simplicity, we decide to focus on the first unitary limit surface $\Sigma_1$, the method
easily generalizes to higher order surfaces $\Sigma_i$ $(i>1)$. Let $\bm{U}$~$\in$~$\Omega$ and $f\in\mathbb{R}^+$
be a factor satisfying
\vspace{-0.2cm}
\begin{equation}
f\bm{U}\in\Sigma_1~.
\vspace{-0.3cm}
\end{equation}
This means scaling $\bm{U}$ by the unique factor $f$ yields a potential on the first unitary limit surface.
While potentials with an empty spectrum must be deepened to obtain a zero-energy bound state, potentials whose
spectrum already contains a dimer with finite binding energy $E<0$ need to be flattened instead. Accordingly,
this behavior is reflected in the following inequalities:
\vspace{-0.2cm}
\begin{equation}
f~~\begin{cases} > 1 & \text{if} \hspace{0.2cm} \bm{U}\in\Omega_0 ~,\\
  = 1 & \text{if} \hspace{0.2cm} \bm{U}\in\Sigma_1 \\ < 1~, & \text{else}~. \end{cases}
\vspace{-0.1cm}
\label{eq-fcases}
\end{equation}
For extremely shallow potentials we observe that $f$ diverges due to the vanishing magnitude of the
potential $\bm{U}$,
\vspace{-0.2cm}
\begin{equation}
\lim_{\bm{U}\to\bm{0}}f = \infty~.
\vspace{-0.1cm}
\end{equation}
Given the factor $f$, we can draw one further conclusion. The radial coordinate of the point $f\bm{U}\in\Sigma_1$
is simply the magnitude $\|f\bm{U}\|$. Then the distance between the potential $\bm{U}$ and the point
$f\bm{U}$ on the first unitary limit surface is given by
\vspace{-0.2cm}
\begin{equation}
x=\|\bm{U}\|-\|f\bm{U}\|=(1-f)\|\bm{U}\|~.
\vspace{-0.2cm}
\end{equation}
Note that $x$ can be negative or vanish due to Eq.~\eqref{eq-fcases}. Although $x$ is not a distance in the
classical sense, it thereby naturally distinguishes between the three cases $\bm{U}$$\in$$\, \Omega_0$,
$\bm{U}$$\in$$\, \Sigma_1$ and $\bm{U}$$\in$~$\Omega\backslash (\Omega_0\cup \Sigma_1)$.

\begin{figure}[t]
\includegraphics[width=0.9\columnwidth]{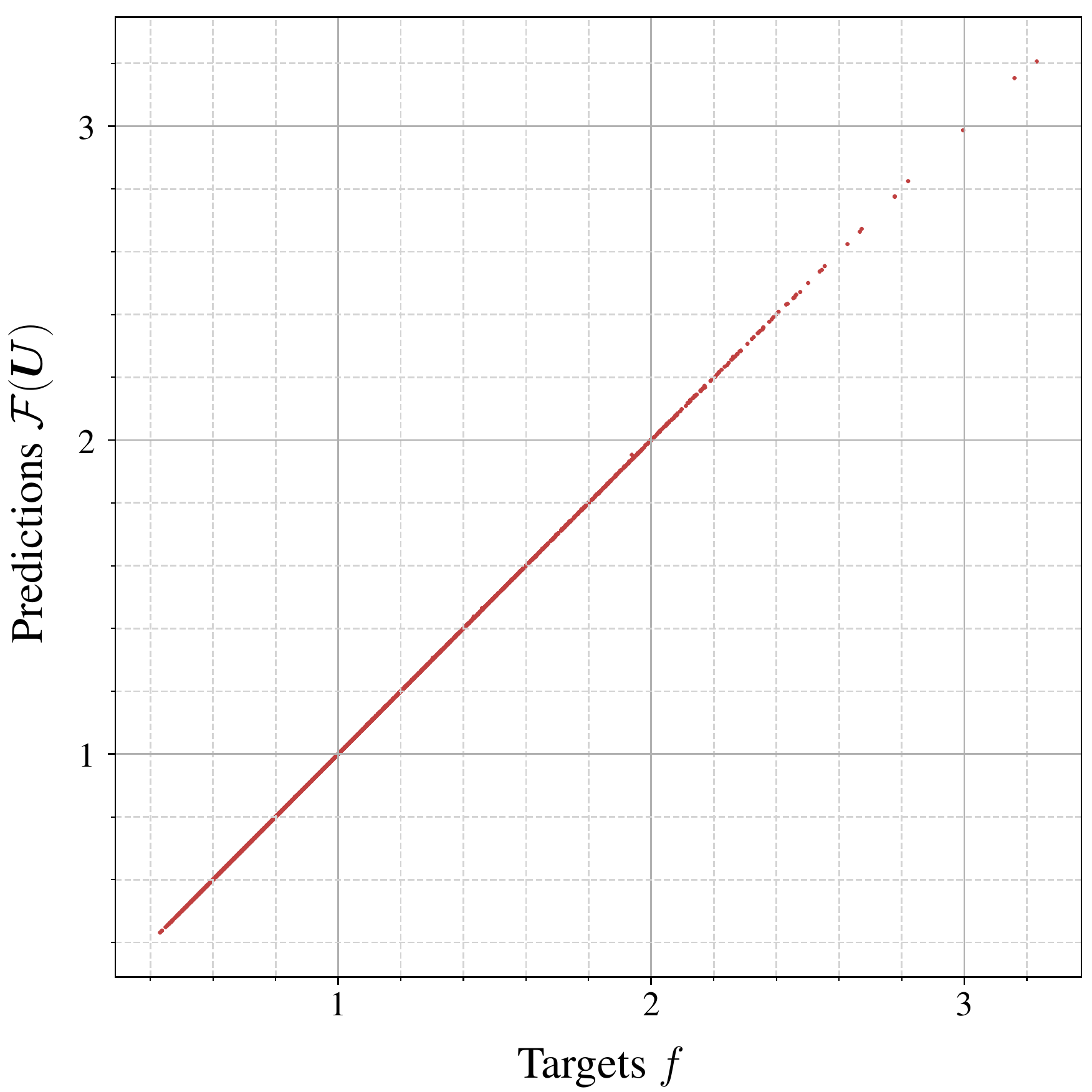}
\caption{Predictions $\mathcal{F}(\bm{U})$ of the scaling factor by the ensemble~$\mathcal{F}$ versus the targets
$f$ for all $(\bm{U},f)\in T_2$. The resulting point cloud is very closely distributed around the bisector,
which indicates an excellent performance of $\mathcal{F}$ on the test set $T_2$.}
\label{fig-1}
\vspace{-2mm}
\end{figure}

\textit{Predicting $f$ with an ensemble of MLPs:}~The factor $f$ seems to be a powerful
quantity for describing the geometry of the unitary limit surface
$\Sigma_1$, the latter is merely the contour for $f=1$. It is a simple task to derive $f$ iteratively by
scaling a given potential $\bm{U}$ until the scattering length flips its sign~\cite{SM}. However, an analytic
relation between $\bm{U}$ and $f$ remains unknown to us. The remedy for this are neural networks that are
trained supervisedly on pairs $(\bm{U},f)\in T_1$ of potentials (inputs) and corresponding factors (targets) in
some training set $T_1$. In this case, neural networks can be understood as maps $\mathcal{F}:\Omega \to
\mathbb{R}$ that additionally depend on numerous internal parameters. The key idea of supervised training
is to adapt the internal parameters iteratively such that the outputs $\mathcal{F}(\bm{U})$ approach
the targets $f$ ever closer. As a result of training, $\mathcal{F}$ approximates the underlying function
$\bm{U}\mapsto f$, such that the factor $f^*\approx\mathcal{F}(\bm{U}^*)$ is predicted with sufficient accuracy
even if the potential $\bm{U}^*\in\Omega$ does not appear in the training set, as long as it resembles the
potentials encountered during training.
%This is also referred to as generalization.
In order to measure the performance of $\mathcal{F}$ on unknown data, one considers a test set $T_2$ containing
previously unknown pairs $(\bm{U}^*,f^*)$ and the mean average percentage error (MAPE) on that data set,
\vspace{-0.2cm}
\begin{equation}
\text{MAPE} = \frac{1}{|T_2|}\sum\limits_{(\bm{U}^*,f^*)\in T_2}\left| \frac{\mathcal{F}(\bm{U}^*)-f^*}{f^*} \right|~.
\vspace{-0.1cm}
\end{equation}
We decide to work with MLPs. These are a widely distributed and very common class of neural networks and
provide an excellent performance for simpler problems. Here, an MLP $\mathcal{F}_i$ with $L$ layers is a composition
\vspace{-0.15cm}
\begin{equation}
\mathcal{F}_i = Y_L \circ \ldots \circ Y_1
\vspace{-0.15cm} 
\end{equation}
of functions $Y_j:~V_{j-1}\to V_{j}$. Usually we have $V_j = \mathbb{R}^{h_j}$. While $Y_1:\Omega\to V_1$ and
$Y_L:V_{L-1}\to \mathbb{R}$ are called the input and output layers, respectively, each intermediate layer
is referred to as a hidden layer. The layer $Y_j$ depends on a weight matrix $W_j~\in~\mathbb{R}^{h_j\times h_{j-1}}$
and a bias $\bm{b}_j\in\mathbb{R}^{h_j}$, both serving as internal parameters, and performs the operation
\vspace{-0.15cm}
\begin{equation}
Y_j(\bm{v})=a_j(W_j\bm{v} + \bm{b}_j)
\vspace{-0.15cm}
\end{equation}
on the vector $\bm{v}\in V_{j-1}$. The function $a_j:\mathbb{R}\to\mathbb{R}$ is called the activation
function of the $j^\text{th}$ layer and is applied component-wise to vectors. Using non-linear
activation functions is crucial in order to make MLPs universal approximators. While output layers are
classically activated via the identity, we activate all other layers via the continuously differentiable
exponential linear unit (CELU) \cite{Barron:2017}, 
\vspace{-0.2cm}
\begin{equation}
\mathrm{CELU}(v)=\mathrm{max}(0,v) + \mathrm{min}(0,\mathrm{exp}(v)-1)~.
\vspace{-0.2cm}
\end{equation}
We use CELU because it is continuously differentiable, has bounded derivatives,
allows positive and negative activations and finally bypasses the
vanishing-gradient-problem, which renders it very useful for deeper architectures. 
In order to achieve precise predictions of the factors $f$, we decide to train an ensemble
of $N_{\mathcal{F}}$~$=$~$100$
MLPs $\mathcal{F}_i$, with each MLP consisting of nine CELU-activated $64$~$\times$~$64$ linear layers and one
output layer. The output of the ensemble is simply the mean of all individual outputs,
\vspace{-0.2cm}
\begin{equation}
\mathcal{F}(\bm{U})=\frac{1}{N_\mathcal{F}}\sum\limits_{i=1}^{N_\mathcal{F}} \mathcal{F}_i(\bm{U})~.
\vspace{-0.2cm}
\end{equation} 
The training and test data sets contain $|T_1|$~$=$~$3$~$\cdot$~$10^4$ and $|T_2|$~$=$~$2.9$~$\cdot$~$10^3$ samples,
respectively, as described in~\cite{SM}. All potentials are discretized with a degree of $d$~$=$~$64$.
Positive and negative scattering
lengths are nearly equally represented in each data set. After $20$ epochs, that is after having scanned through
the training set for the $20^\text{th}$ time, the training procedure is terminated and the resulting MAPE
of the ensemble $\mathcal{F}$ turns out as $0.028\,\%$. When plotting predictions versus
targets, this implies a thin point cloud that is closely distributed around the bisector as can be seen
in Fig.~\ref{fig-1}. We therefore conclude that $\mathcal{F}$ returns very precise predictions on $f$.
\begin{figure}[t]
\includegraphics[width=0.9\columnwidth]{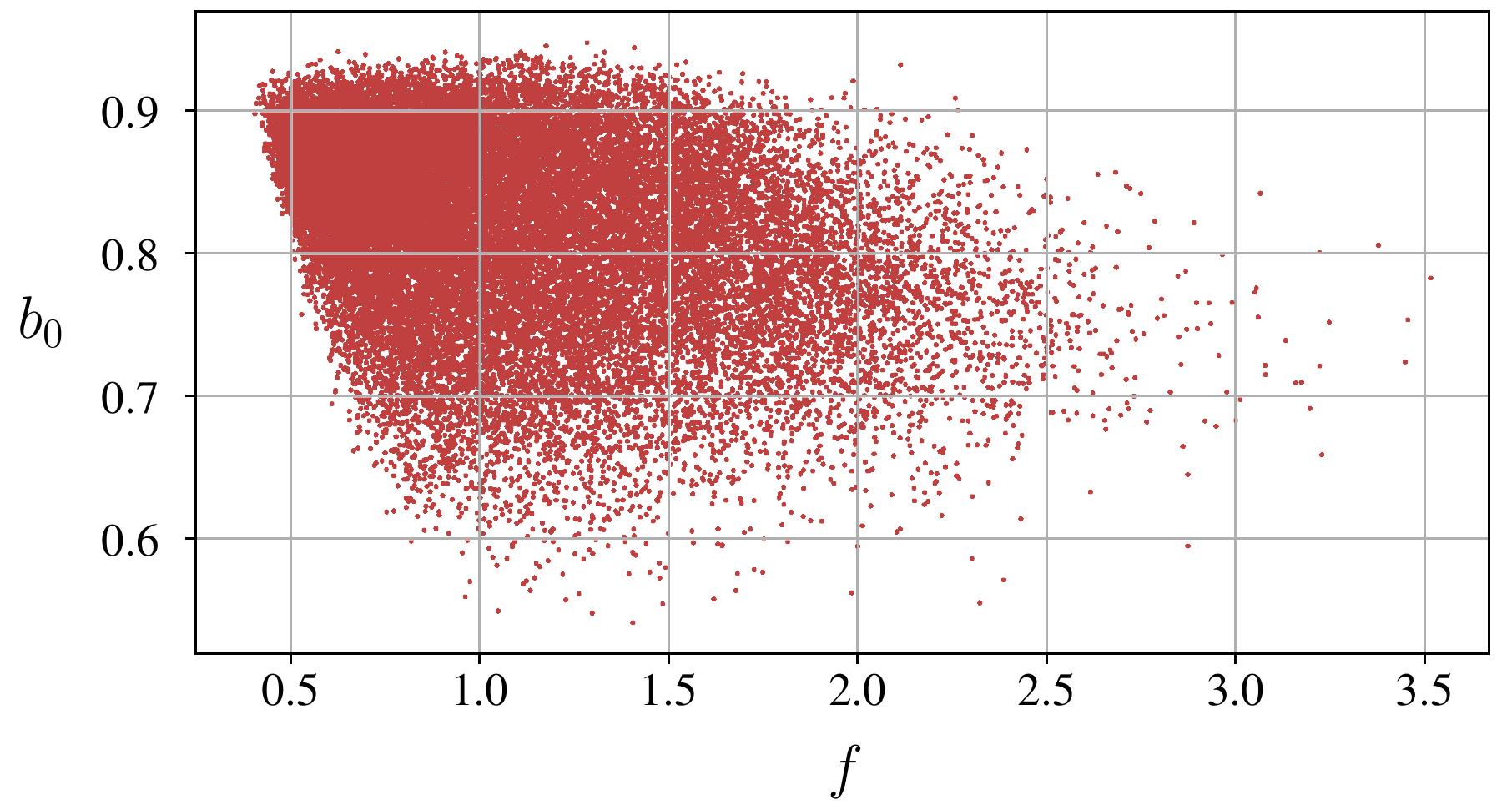}
\caption{$b_0$ versus the corresponding factors $f$ for all potentials $\bm{U}$ in the training set $T_1$.
Note that $b_0$ is restricted to a small interval. The width of the point cloud suggests that there is no
one-to-one relation between $b_0$ and $f$.}
\label{fig-2}
\vspace{-2mm}
\end{figure}

\textit{Predicting scattering lengths in vicinity of $\mathit{\Sigma_1}$:}~As shown in the previous section,
training an ensemble of MLPs to reproduce the factors $f$ for given potentials $\bm{U}\in\Omega$ is a
numerically simple approach for investigating the unitary limit surface $\Sigma_1$. Our key motivation is
to predict scattering lengths in the vicinity of unitary limit surfaces by using neural networks. However, the
unitary limit itself poses a major obstacle to common neural networks and training algorithms: Reproducing
the movable singularity for $a_0$ on $\Sigma_1$ imposes severe restrictions on the MLP architecture and
renders the training steps unstable. Thus we must pursue alternative approaches. The idea of the approach we
opt for is to express scattering lengths in terms of regular quantities, that each can be easily predicted by
MLPs. Therefore we first consider the quantity
\vspace{-0.1cm}
\begin{equation}
b_0 = a_0(1-f) = \frac{a_0 x}{\|\bm{U}\|}~.
\label{eq-bafrelation}
\vspace{-0.1cm}
\end{equation}
As shown in Fig.~\ref{fig-2}, $b_0$ is finite and restricted to a small interval for all potentials in the training
set. We therefore expect training MLPs to predict $b_0$ to be a numerically simple task. This again suggests to
train an ensemble $\mathcal{B}$ of $N_\mathcal{B}=100$ MLPs~$\mathcal{B}_i$,
\vspace{-0.55cm}
\begin{equation}
\mathcal{B}(\bm{U})=\frac{1}{N_\mathcal{B}}\sum\limits_{i=1}^{N_\mathcal{B}} \mathcal{B}_i(\bm{U})~.
\vspace{-0.15cm}
\end{equation}
While the members $\mathcal{B}_i$ only consist of five CELU-activated $64$~$\times$~$64$ linear layers and one
output layer, the rest coincides with the training procedure of the ensemble $\mathcal{F}$ as presented in
the previous section. The resulting MAPE of the ensemble $\mathcal{B}$ turns out as $0.017\,\%$,
which indicates that $\mathcal{B}$ approximates the relation $\bm{U}\mapsto b_0$ very well around $\Sigma_1$.

\begin{figure}[t]
\includegraphics[width=0.9\columnwidth]{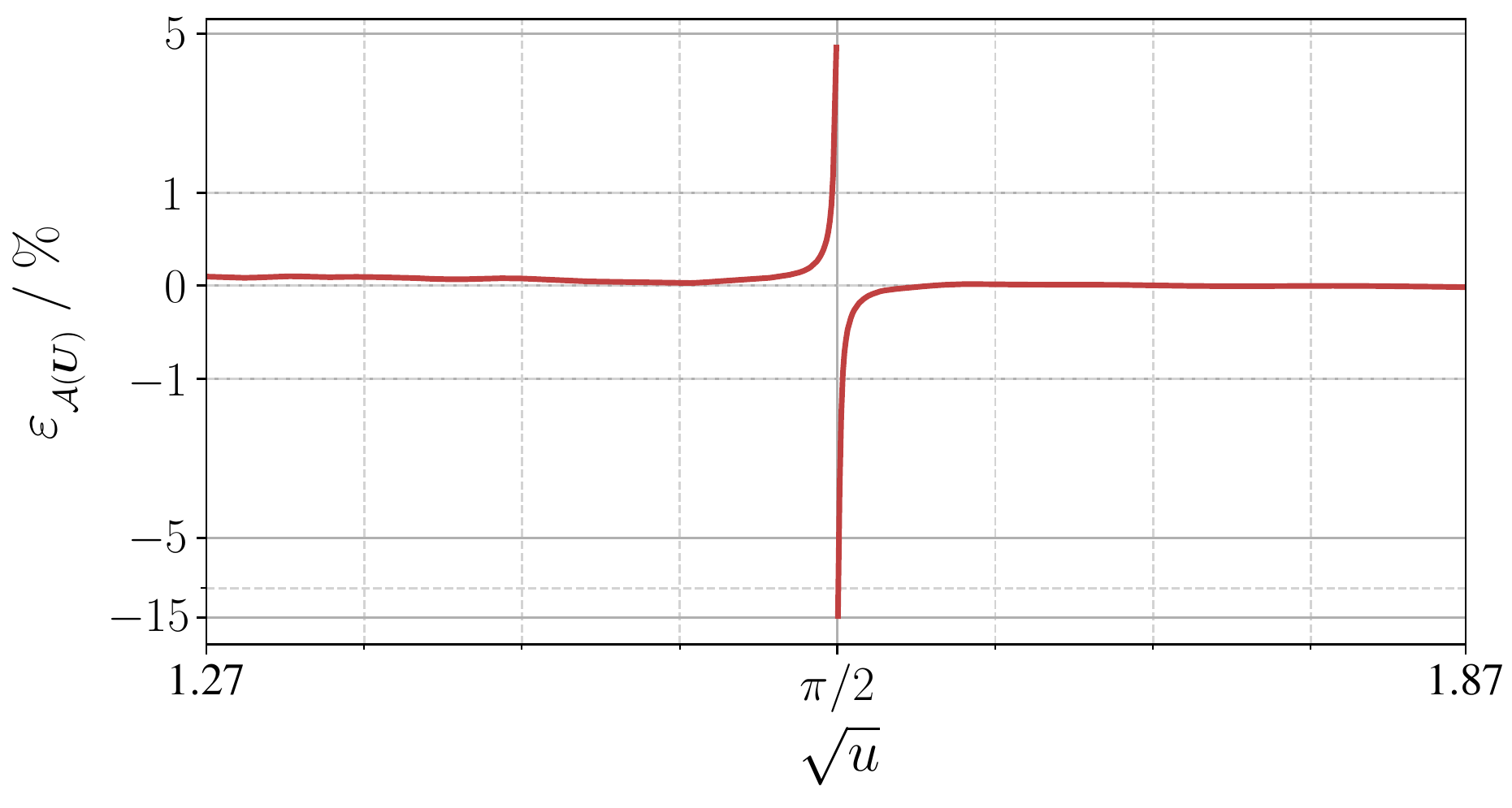}
\caption{Relative errors $\varepsilon_{{}_{\mathcal{A}(\bm{U})}}$ of predicted scattering lengths $\mathcal{A}(\bm{U})$
for potential wells $\mathcal{U}$ with depths $u$. The $\varepsilon_{{}_{\mathcal{A}(\bm{U})}}$ only take values between $1\%$
and $15\%$ in close vicinity of the unitary limit at $u~=~\pi^2/4$. Otherwise, the relative errors become
negligibly small.}
\label{fig-4}
\end{figure}

\begin{figure}[t]
\includegraphics[width=0.9\columnwidth]{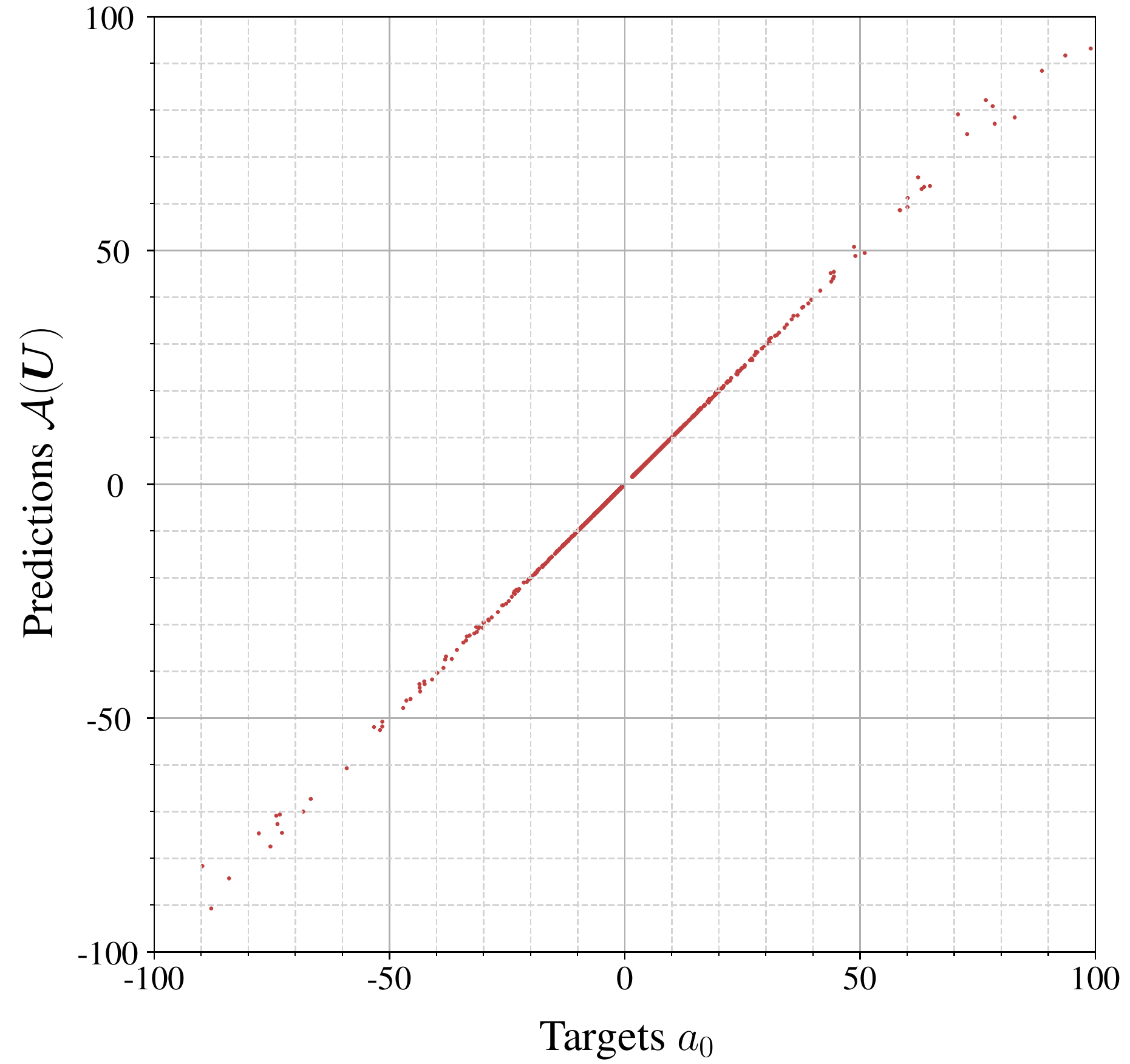}
\caption{Predictions $\mathcal{A}(\bm{U})$ of scattering lengths by the quotient $\mathcal{A}$ versus the targets
$a_0$. The point cloud becomes broader for unnaturally large scattering lengths. Nonetheless it is still distributed
sufficiently close around the bisector, which indicates that $\mathcal{A}$ generalizes well and reproduces the
correct behavior of $a_0$ around $\Sigma_1$.}
\label{fig-5}
\vspace{-2mm}
\end{figure}

Due to Eq.~\eqref{eq-bafrelation}, scattering lengths can be expressed in terms of $b_0$ and $f$. Having
trained the ensembles $\mathcal{B}$ and $\mathcal{F}$ to predict these two quantities precisely, we expect
the quotient
\vspace{-0.1cm}
\begin{equation}
\mathcal{A}(\bm{U})=\frac{\mathcal{B}(\bm{U})}{1-\mathcal{F}(\bm{U})}
\vspace{-0.1cm}
\end{equation}
to provide a good approximation of $a_0$ for potentials $\bm{U}$$\in$$\, \Omega$ in vicinity of $\Sigma_1$.
However, note that outputs $\mathcal{A}(\bm{U})$ for potentials $\bm{U}$ in the unitary limit $f\to 1$ are
very sensitive to $\mathcal{F}(\bm{U})$. In this regime, even the smallest errors may cause a large deviation
from the target values and thereby corrupt the accuracy of $\mathcal{A}$. Let us therefore consider the
relative errors of predicted scattering lengths,
\vspace{-0.2cm}
\begin{equation}
\varepsilon_{{}_{\mathcal{A}(\bm{U})}}=\frac{\mathcal{A}(\bm{U})-a_0}{a_0}~,
\vspace{-0.2cm}
\end{equation}
for potential wells $\bm{U}=u\begin{pmatrix} 1, & \ldots, & 1 \end{pmatrix}^T$ with depths $u$. In Fig.~\ref{fig-4}
we observe significantly larger relative errors in a small interval around the unitary limit at $u=\pi^2/4$.
Nonetheless, the quotient $\mathcal{A}$ reproduces the behavior of $a_0$ sufficiently well. 
We can also convince ourselves of this for more general potentials by inspecting the prediction-vs-target plot
in Fig.~\ref{fig-5}: Although we notice a broadening of the point cloud for unnaturally large scattering lengths,
the point cloud itself remains clearly distributed around the bisector. Finally, the resulting MAPE of
$0.41\,\%$ indicates an overall good performance of $\mathcal{A}$ on the test set $T_2$.
%\newpage

\textit{Discussion and outlook:}~The unitary limit $a_0\to\infty$ is realized by movable singularities
in potential space $\Omega$, each
corresponding to a hypersurface $\Sigma_i\subset\Omega$ that we refer to as the $i^\text{th}$ unitary limit
surface. This formalism not only lets one understand the unitary limit in a geometric manner, but also
introduces new quantities $f$ and $b_0$. These are regular in the unitary limit and provide an alternative
parameterization of low-energy scattering processes. As such, they suffice to derive the S-wave scattering length
$a_0$. By training ensembles of multilayer perceptrons in order to predict $f$ and $b_0$, respectively, we
therefore successfully establish a machine learning based description for unnatural as well as natural scattering
lengths.

Note that this is by far not the only reasonable approach towards the unitary limit. Another, even simpler
approach would have been to train networks to predict the inverse scattering length $1/a_0$, which is obviously
regular in the unitary limit. Considering the inverse prediction afterwards would provide a good estimate on
unnatural scattering lengths, too. However, we have decided to stay in the developed formalism of unitary
limit surfaces and to use the geometric quantities $f$ and $b_0$ to describe the unitary limit for the sake
of interpretability. Concerning interpretability, it is important to note that both trained ensembles still
need to be considered as ``black boxes'', since we have not interpreted how inputs are processed physically,
yet. An appropriate approach to mention here is the Taylor decomposition of both ensembles for expansion
points on $\Sigma_1$~\cite{SM}. This also provides additional geometric insights
like normal vectors on the \linebreak unitary limit surface.

\pagebreak

Of course, this approach is also suitable for describing unitary limit surfaces of higher order.
We can also think of a simultaneous treatment of several unitary limit surfaces. Also, it can be generalized
to arbitrary effective range expansion parameters that produce other movable singularities in potential space. 
More generally, its scope is  not limited to scattering problems, but can be utilized in  systems that exhibit an
unnaturally large scale.

\vspace{0.4cm}

{\it
We thank Bernard Metsch for useful comments.  
We acknowledge partial financial support from the Deutsche Forschungsgemeinschaft (TRR 110,
``Symmetries and the Emergence of Structure in QCD''),
Further support was provided by the Chinese Academy of Sciences
(CAS) President's International Fellowship Initiative
(PIFI) (grant no. 2018DM0034) and by VolkswagenStiftung (grant no. 93562).
%The computational resources were provided by the Julich Supercomputing
%Centre at Forschungszentrum J\"ulich, Oak Ridge Leadership Computing
%Facility, RWTH Aachen, and Michigan State University.
}

\newpage

\section*{Supplemental Material}

\textit{Preparation of data sets:}~In order to make potentials processable for neural networks, they have to be
discretized. We associate the discretized potential $\bm{U}\in\Omega$ with the piecewise constant step potential
\begin{equation}
U(r)=\begin{cases} U_1 &\text{if}\ \hspace{0.45cm} r_0\leq r< r_1~, \\ &\vdots \\ U_{n} &\text{if}\
\hspace{0.45cm} r_{n-1}\leq r< r_{n}~, \\ &\vdots \\ 0 &\text{if}\ \hspace{0.45cm} r_{d}\leq r< r_{d+1}~, \end{cases}
\label{eq-dt-pr-3}
\end{equation}
with the transition points $r_n=n/d$ for $n=0,\ldots,d$ and $r_{d+1}=\infty$. Accordingly, the $l^\text{th}$ partial
wave is defined piecewise as well: Between the transition points $r_{n-1}$ and $r_n$ it is given as a linear
combination of spherical Bessel and Neumann functions,
\begin{equation}
\phi_{n}^{(l,k)}(r)=A_{l,n}(k)j_l(k_nr)-B_{l,n}(k)n_l(k_nr)~, \label{eq-partialwave}
\end{equation}
with the kinetic energy
\begin{equation}
k_n=\Theta_\mathrm{Re}(k)\sqrt{k^2+U_n\vphantom{\frac{a}{a}}}~.
\end{equation}
Here we introduce the factor
\begin{equation}
\Theta_\mathrm{Re}(k)=\begin{cases} +1~, &\text{if}\ \mathrm{Re}(k)\geq 0 \\ -1~, &\text{if}\ \mathrm{Re}(k)< 0
\end{cases}
\end{equation}
to conserve the sign of $k$ on the complex plane, that is $k_n$~$\to$~$k$, if $U_n$ vanishes. The parameters
$A_{l,d+1}(k)$ and $B_{l,d+1}(k)$ completely determine the effective range function $K_l(k)~=~k^{2l+1}~\cot \delta_l(k)$
due to their asymptotic behavior
\begin{align}
A_{l,d+1}(k)&=\mathrm{e}^{i\delta_l(k)}\cos\delta_l(k),\\ B_{l,d+1}(k)&=\mathrm{e}^{i\delta_l(k)}\sin\delta_l(k)~.
\end{align}
Instead of solving the Schr\"odinger equation for the step potential $U(r)$, we apply the transfer matrix method
\cite{Jonsson:1990} to derive $A_{l,d+1}(k)$ and $B_{l,d+1}(k)$. Due to the smoothness of the partial wave
$\phi^{(l,k)}$ at each transition point $r_n$, this method allows us to relate $A_{l,d+1}(k)$ and $B_{l,d+1}(k)$
to the initial parameters $A_{l,1}(k)$ and $B_{l,1}(k)$ via a product of transfer matrices $M_{l,n}(k)$. To arrive
at a representation of these transfer matrices, we split up the mentioned smoothness condition into two separate
conditions for continuity,
\begin{equation}\phi_{n+1}^{(l,k)}(r_{n})=\phi_{n}^{(l,k)}(r_{n})\label{eq-continuity}\end{equation}
and differentiability,
\begin{equation}
\frac{\mathrm{d}}{\mathrm{d}r}\phi_{n+1}^{(l,k)}(r)\bigg|_{r=r_{n}}
= \frac{\mathrm{d}}{\mathrm{d}r}\phi_{n}^{(l,k)}(r)\bigg|_{r=r_{n}}
\label{eq-diffability}
\end{equation}
at each transition point $r_{n}$. Using Eq. \eqref{eq-partialwave}, we can combine both conditions \eqref{eq-continuity} and \eqref{eq-diffability} to a vector equation, that connects neighboring coefficients with each other:
\begin{align}
\underbrace{\begin{pmatrix} j_l(k_{n+1}r_{n}) & -n_l(k_{n+1}r_{n}) \\ k_{n+1}j_l^\prime(k_{n+1}r_{n}) &
-k_{n+1}n_l^\prime(k_{n+1}r_{n}) \end{pmatrix}}_{=m_{l,n+1}(r_n,k)}\begin{pmatrix} A_{l,n+1}(k) \\ B_{l,n+1}(k)\end{pmatrix}
\notag \\ =\underbrace{\begin{pmatrix} j_l(k_nr_{n}) & -n_l(k_nr_{n}) \\ k_nj_l^\prime(k_nr_{n}) & -k_nn_l^\prime(k_nr_{n})
\end{pmatrix}}_{=m_{l,n}(r_n,k)}\begin{pmatrix} A_{l,n}(k) \\ B_{l,n}(k)\end{pmatrix}~. \label{eq-trf-1}
\end{align}
Multiplying Eq.~\eqref{eq-trf-1} with $m_{l,n+1}^{-1}(r_n,k)$ from the left yields
\begin{equation}
\begin{pmatrix} A_{l,n+1}(k) \\ B_{l,n+1}(k)\end{pmatrix} = M_{l,n}(k)
\begin{pmatrix} A_{l,n}(k) \\ B_{l,n}(k)\end{pmatrix}, \label{eq-trf-2}
\end{equation}
which defines the $n^\text{th}$ transfer matrix
\begin{equation}
M_{l,n}(k)=m_{l,n+1}^{-1}(r_n,k)m_{l,n}^{}(r_n,k)~.\label{eq-trf-3}
\end{equation}
Therefore, $A_{l,d+1}(k)$ and $B_{l,d+1}(k)$ are determined by the choice of $A_{l,1}(k)$ and $B_{l,1}(k)$,
which requires us to define two boundary conditions. Due to the singularity of $n_l$ in the origin,
the spherical Neumann contribution in the first layer must vanish and therefore $B_{l,1}(k)=0$. The choice
of $A_{l,1}(k)$ may alter the normalization of the wave function. However, since we only consider ratios
of $A_{l,d+1}(k)$ and $B_{l,d+1}(k)$, we may opt for $A_{l,1}(k)=1$, which corresponds to
\begin{equation}
\phi_{1}^{(l,k)}(r)=j_l^{}(k_1r)~.
\end{equation}
Finally, applying all transfer matrices successively to the initial parameters yields
%%due to therecursion in Eq.~\eqref{eq-trf-1}
\begin{equation}
\begin{pmatrix} A_{l,d+1}(k) \\ B_{l,d+1}(k)\end{pmatrix} = \left(\prod\limits_{n=1}^{d}M_{l,n}(k)\right) \ \begin{pmatrix} 1 \\ 0 \end{pmatrix}~. \label{eq-trf-4}
\end{equation}

The most general way to derive any effective range expansion parameter $Q_l^{(j)}(\varkappa)$ for arbitrary
expansion points $\varkappa\in\mathbb{C}$ in the complex momentum plane is a contour integration along a
circular contour $\gamma$ with radius $\kappa_\gamma$ around $\varkappa$. Applying Cauchy's integral
theorem then yields
\begin{align}
Q_l^{(j)}(\varkappa)&=\frac{1}{2\pi i}\oint_\gamma \! \mathrm{d}k \,  \frac{k^{2l+1}}{(k-\varkappa)^{j+1}}
\frac{A_{l,d+1}(k)}{B_{l,d+1}(k)}~.
\end{align}
We approximate this integral numerically over $N$ grid points
\begin{equation}
k_q = \varkappa+\kappa_\gamma \mathrm{e}^{iq{2\pi}/{N}},\hspace{1cm} q=0,\ldots,N-1~. 
\end{equation}
Smaller contour radii $\kappa_\gamma$ and larger $N$ thereby produce finer grids and decrease the approximation
error. This way of calculating $Q_l^{(j)}(\varkappa)$ requires in total $d$~$\times$~$N$ transfer matrices.
The numerical integration provides
\begin{equation}
Q_l^{(j)}(\varkappa)\approx\frac{1}{N(\kappa_\gamma)^j}\sum\limits_{q=0}^{N-1}(k_q)^{2l+1}
\mathrm{e}^{-iq\frac{2\pi}{N}}\frac{A_{l,d+1}(k_q)}{B_{l,d+1}(k_q)}~.
\label{eq-numint}
\end{equation}
Despite the generality of Eq.~\eqref{eq-numint}, we restrict this analysis to $S$-wave scattering lengths
$a_0$~$=$~$-1/Q_0^{(0)}(0)$, since these dominate low-energy scattering processes.

While generating the training and test sets, we must ensure that there are no overrepresented potential shapes
among the respective data set. To maintain complexity, this suggests generating potentials with randomized
components $U_n$. An intuitive approach therefore is to produce them via Gaussian random walks: Given $d$
normally distributed random variables $X_1,\ldots,X_{d}$,
\begin{equation}
X_i \sim \begin{cases} \mathcal{N}(0,\mathrm{ISF}) &\text{if}\ i=1~,\\
\mathcal{N}(0,\mathrm{SF}) &\text{else}~, \end{cases}
\label{eq-rw-1}
\end{equation}
where $\mathcal{N}(\mu,\sigma)$ describes a normal distribution with mean $\mu$ and standard deviation $\sigma$,
the distribution of the $n^\text{th}$ potential step $U_n$ can be described by the magnitude of the sum over
all previous steps $X_i$,
\begin{equation}
U_n=\left|\sum\limits_{i=1}^n X_i\right|~. \label{eq-rw-2}
\end{equation}
Note that, while all steps $X_i$ in Eq.~\eqref{eq-rw-1} have zero mean, the standard deviation of the first step,
which we denote by the initial step factor $\mathrm{ISF}$, may differ from the standard deviation of all
other steps, that we refer to as the step factor $\mathrm{SF}$. This allows us to roughly control the shapes and depths of all
potentials in the data set. Choosing $\mathrm{ISF}\gg \mathrm{SF}$ results more likely in potentials that
resemble a potential well and expectedly yield similar scattering lengths. In contrast to that,
$\mathrm{SF}\gg \mathrm{ISF}$ produces strongly oscillating potentials.
We decide to choose the middle course $\mathrm{ISF}=1$ and $\mathrm{SF}=0.75$ 
(this is the case $\mathrm{SF}\approx \mathrm{ISF}$) for two reasons: For one, from the perspective of depths, 
the corresponding Gaussian random walk is capable of generating 
potentials around the first unitary limit surface $\Sigma_1$.
For another, this choice of step factors causes the data set to cover a wide range of shapes from 
dominantly potential wells to more oscillatory potentials, which is an important requirement for generalization.
This way, we generate $10^5$ potentials for the training set and $10^4$ potentials for the test set,
To avoid overfitting to a certain potential depth, this needs to be followed by a rigorous
downsampling procedure. For this, we use the average depth
\begin{equation}
u = \frac{1}{d}\sum\limits_{n=1}^d U_n
\end{equation}
\begin{figure}[t]
\includegraphics[width=\columnwidth]{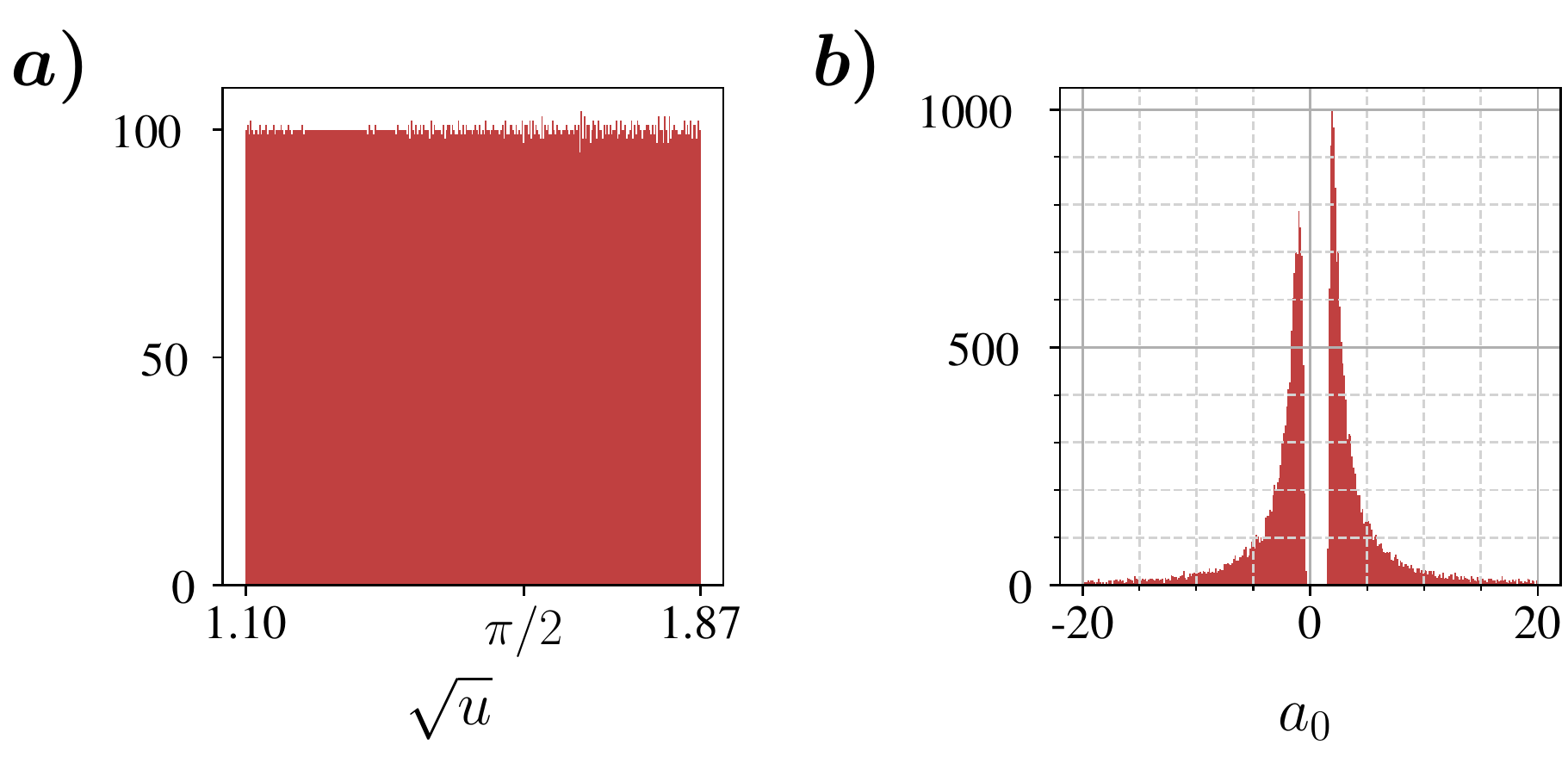}
\caption{$a):$ Distribution of the square root $\sqrt{u}$ of the average depth over the training set. By
  construction, this distribution is uniform. $b):$ Bimodal distribution of the scattering length $a_0$
  over the training set. Note that extremely large scattering lengths are not displayed in this histogram.}
\vspace{-2mm}
\label{figSM-1}
\end{figure}\noindent
as a measure. Uniformizing the training and test set with respect to $\sqrt{u}$ on the interval
$[1.10,1.87]$ by randomly omitting potentials with overrepresented depths finally yields a training set
$T_1$, see Fig.~\ref{figSM-1}, and test set $T_2$ that contain $3\times 10^4$ and $2.9\times 10^3$ potentials,
respectively. Scattering lengths are then derived using the numerical contour integration in Eq.~\eqref{eq-numint}
with $N=100$ grid points and a contour radius of $\kappa_\gamma=0.1$. The derivation of the factors $f$ is more
involved: If the scattering length is negative (positive), the potential is iteratively scaled with the factor
$s=2$ ($s=1/2$), until its scattering length changes its sign. Let us assume the potential has been scaled $t$
times this way. Then we can specify the interval where we expect to find $f$ in as $(2^{t-1},2^t]$ or as $[2^{-t},2^{1-t})$,
respectively. Cubically interpolating $1/a_0$ on that interval using $25$ equidistant values and searching for its
zero finally yields the desired factor $f$. 

\medskip

\textit{Training by gradient descent:}~Given a data set $D$~$\subseteq$~$\Omega$~$\times$~$\mathbb{R}^n$, there
are several ways to measure the performance of a neural network $\mathcal{N}:\Omega\to\mathbb{R}^n$ on $D$.
For this we have already introduced the MAPE that we have derived for the test set $D=T_2$ after training.
Lower MAPEs are thereby associated with better performances. Such a function $L:\Gamma\to\mathbb{R}^+$ that
maps a neural network to a non-negative, real number is called a loss function. The weight space $\Gamma$
is the configuration space of the used neural network architecture and as such it is spanned by all
internal parameters (e.g. all weights and biases of an MLP). Therefore, we can understand all neural networks
$\mathcal{N}\in\Gamma$ of the given architecture as points in weight space. The goal all training algorithms
have in common is to find the global minimum of a given loss function in weight space. It is important to note
that loss functions become highly non-convex for larger data sets and deeper and more sophisticated
architectures. As a consequence, training usually reduces to finding a well performing local minimum.

%In contrast to the MAPE, there are other loss functions that play an important role during training, which will be explained below in more detail. For regression the two most commonly used loss functions are the mean-squared-error loss~$\text{(MSE-Loss)}$,
%\begin{equation}
%L_\text{MSE}(\bm{t},\mathcal{N}(\bm{U}))=\frac{1}{N}(\bm{t}-\mathcal{N}(\bm{U}))^2 \label{eq-mse}
%\end{equation}
%and the mean-absolute-error loss (MAE or L1-Loss),
%\begin{equation}
%L_\text{L1}(\bm{t},\mathcal{N}(\bm{U}))=\frac{1}{N}\|\bm{t}-\mathcal{N}(\bm{U})\|. \label{eq-l1}
%\end{equation}
%While the losses in Eqs. \eqref{eq-mse} and \eqref{eq-l1} only allow for a single sample $(\bm{U},\bm{t})\in\Omega$~$\times$~$\mathbb{R}^n$, we can use any loss function $L$ to construct batch losses for arbitrary datasets $D$,
%\begin{equation}
%L_D(\mathcal{N}) = \frac{1}{|D|} \sum\limits_{(\bm{U},\bm{t})\in D}L(\bm{t},\bm{N}(\bm{U})).
%\end{equation}

A prominent family of training algorithms are gradient descent techniques. These are iterative with each
iteration corresponding to a step the network takes in weight space. The direction of the steepest
loss descent at the current position $\mathcal{N}\in\Gamma$ is given by the negative gradient of
$L(\bm{t},\mathcal{N}(\bm{U}))$. Updating internal parameters along this direction is the
name-giving feature of gradient descent techniques. This suggests the update rule
\begin{equation}
p \longleftarrow p-\eta\frac{\partial L}{\partial p}(\bm{t},\mathcal{N}(\bm{U}))
\end{equation}
for each internal parameter $p$, and by the left arrow $a~\longleftarrow~b$ we denote the assignment
'{\verb+a=b+}' as used in computer programming. Accordingly, the entire training procedure corresponds to
a path in weight space. The granularity of that path is controlled by the learning rate $\eta$: Smaller
learning rates cause a smoother path but a slower approach towards local minima and vice versa. In any case,
training only for one epoch, that is scanning only once through the training set, usually does not suffice
to arrive near any satisfactory minima. A typical training procedure consists of several epochs.

Usually, the order of training samples $(\bm{U},\bm{t})$ is randomized to achieve a
faster learning progress and to make training more robust to badly performing local minima.
Therefore, this technique is also called stochastic gradient descent. Important alternatives to mention are
mini-batch gradient descent  and batch gradient descent, where update steps are not taken with respect
to the loss $L(\bm{t},\mathcal{N}(\bm{U}))$ of a single sample, but to the batch~loss 
\begin{equation}
L_D(\mathcal{N}) = \frac{1}{|D|} \sum\limits_{(\bm{U},\bm{t})\in D}L(\bm{t},\mathcal{N}(\bm{U}))
\end{equation}
of randomly selected subsets $D$ of the training set $T_1$ with the batch size $|D|=B$ or the entire
training set itself, respectively. There are more advanced gradient descent techniques like Adam and Adamax
\cite{Kingma:2017} that introduce a dependence on previous updates and adapted learning rates. It is
particularly recommended to use these techniques when dealing with large amounts of data and high-dimensional
weight spaces.

For training the members $\mathcal{F}_i$ and $\mathcal{B}_i$ of both ensembles $\mathcal{F}$ and $\mathcal{B}$,
we apply the same training procedure using the machine learning framework provided by PyTorch~\cite{Paszke:2019}: Weights and biases are initialized via the He-initialization~\cite{He:2015}.
We use the Adamax optimizer with the batch size $B=10$ to minimize the L1-Loss,
\begin{equation}
L_\text{L1}(\bm{t},\mathcal{N}(\bm{U}))=\frac{1}{n}\|\bm{t}-\mathcal{N}(\bm{U})\|,
\end{equation}
over $20$ epochs. Here, we apply an exponentially decaying learning rate schedule, that is $\eta_t = 0.01\ \exp (-t/2)$ is epoch dependent. In this case, the decreasing learning rates allow a much closer and more stable approach towards local minima.

\medskip

\begin{figure}[t]
\includegraphics[width=\columnwidth]{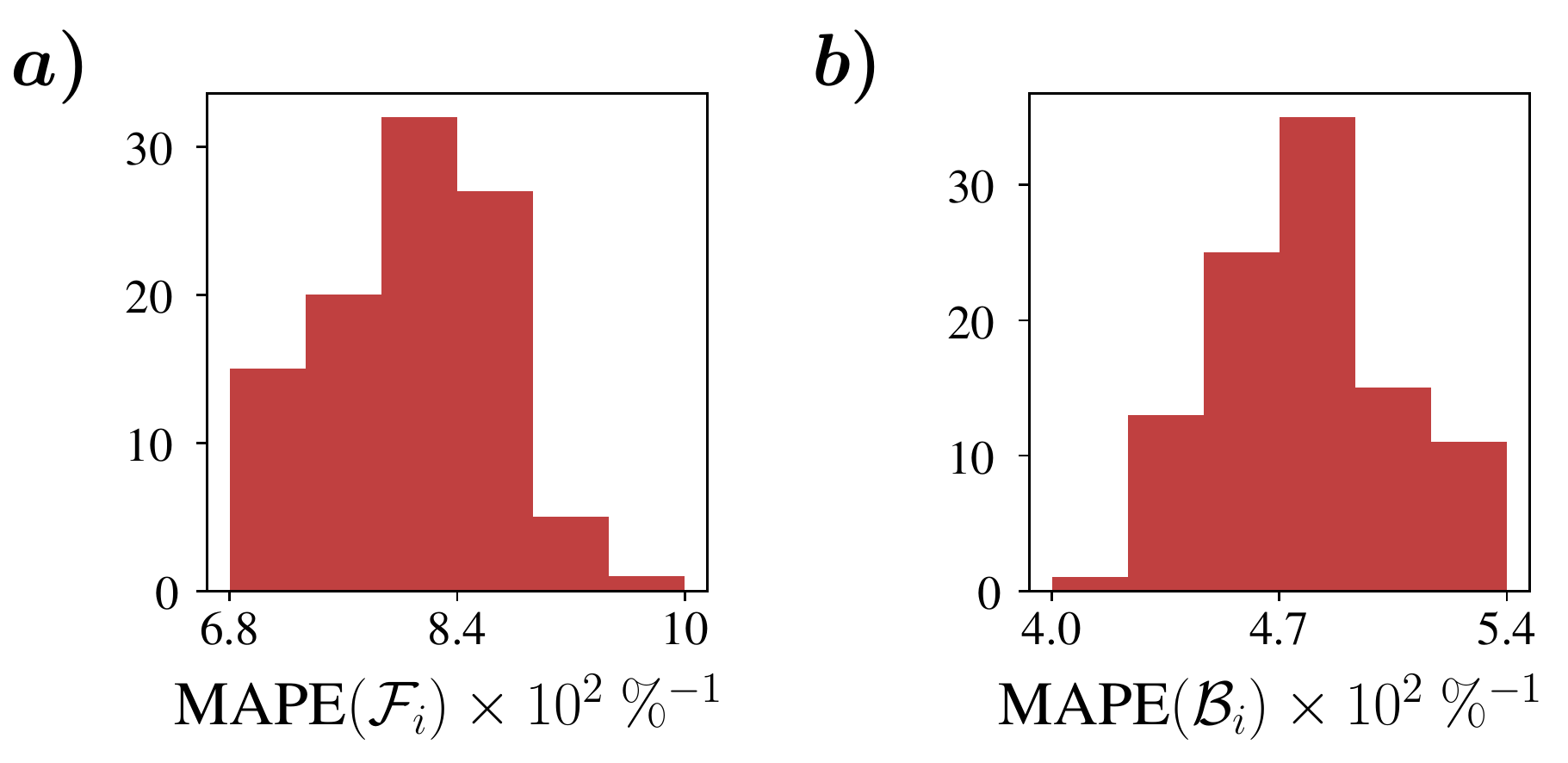}
\caption{$a):$ Individual MAPEs of the members $\mathcal{F}_i$ of the ensemble $\mathcal{F}$. $b):$ Individual MAPEs of the members $\mathcal{B}_i$ of the ensemble $\mathcal{B}$.}
\label{figSM-2}
\end{figure}

\textit{Neural network ensembles:}~Each ensemble $\mathcal{F}$ and $\mathcal{B}$ consists of $100$ MLPs. In
the above framework, an ensemble can be understood as a point cloud in weight space. We can assume that each
of these members is close to a well performing local minimum or even to the global minimum. There are several
techniques to let the ensemble vote on one, even more precise prediction. Here, we choose to take the average
of all individual predictions. We observe a significant increase in accuracy (by almost a factor of two)
by comparing the resulting MAPEs of $\mathcal{F}$ and $\mathcal{B}$ with the individual MAPEs of the members,
shown in Figs.~\ref{figSM-2}a) and \ref{figSM-2}b).
\noindent
\begin{figure}[t]
\includegraphics[width=\columnwidth]{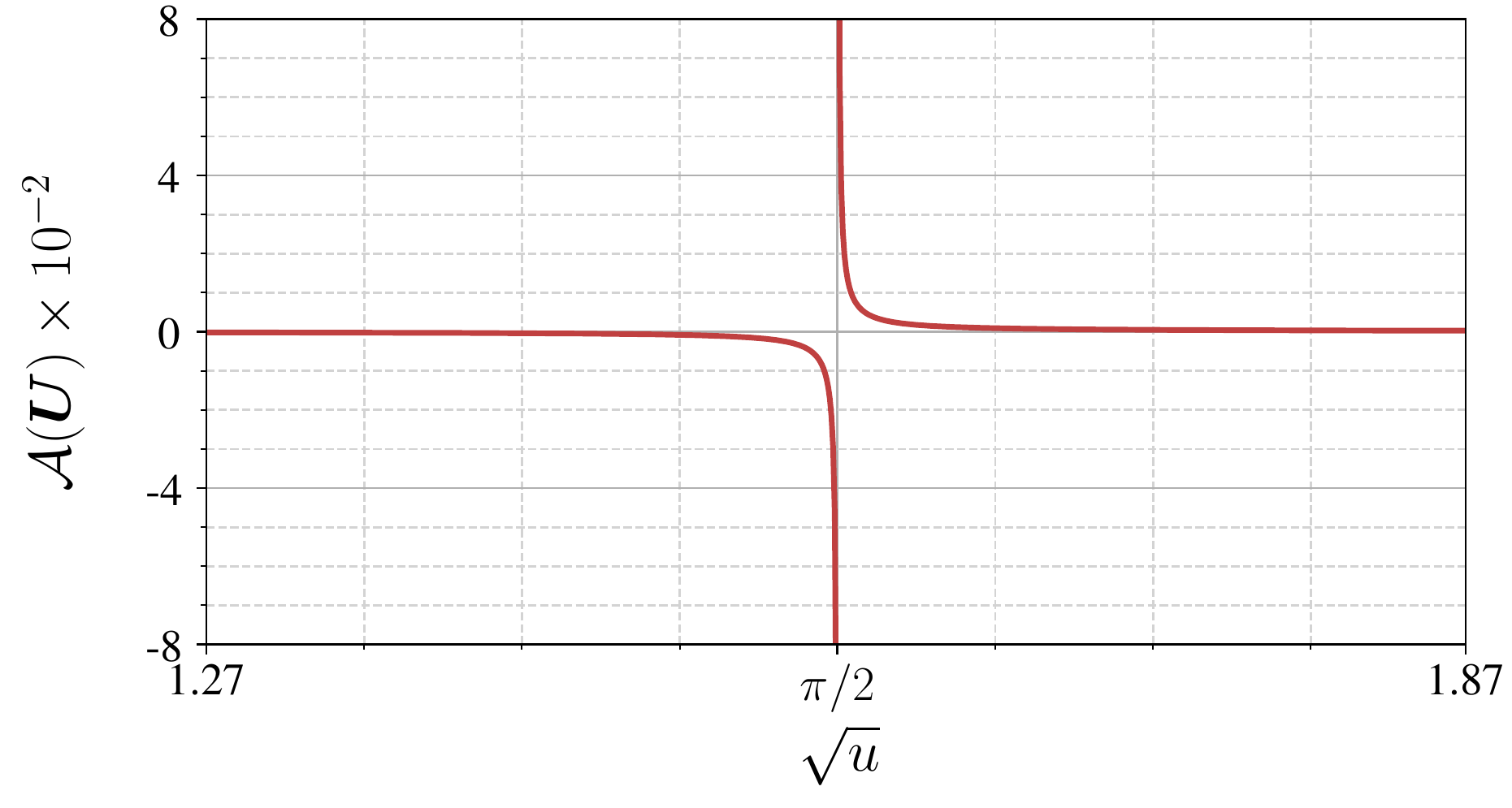}
\caption{Predictions of natural and unnatural scattering lengths for potential wells $\bm{U}=u\begin{pmatrix}1,\ldots,1\end{pmatrix}$ by the quotient $\mathcal{A}$. Note that the shown graph reproduces the expected behavior $a_0$~$=$~$1$~$-$~$\tan\sqrt{u}/\sqrt{u}$ very well.}
\vspace{-2mm}
\label{figSM-3}
\end{figure}\noindent

\medskip

\textit{Taylor expansion to restore interpretability:}~Considering the quotient
$\mathcal{A}(\bm{U})$$=$$\mathcal{B}(\bm{U})$$/$$(1$$-$$\mathcal{F}(\bm{U}))$, we can make reliable
predictions on natural and unnatural scattering lengths, c.f.~Fig.~\ref{figSM-3}. We have established
a geometrical understanding of the quantities $f$ and $b_0$, predicted by $\mathcal{F}$ and $\mathcal{B}$,
respectively. However, since both ensembles are ``black boxes'', their outputs and the outputs of
$\mathcal{A}$ are no longer interpretable beyond that level. One way to restore interpretability is
to consider the Taylor expansion with respect to an appropriate expansion point. In the following, we
demonstrate this for the ensemble $\mathcal{F}$ on the first unitary limit surface: Since $\mathcal{F}$ is
regular in the expansion point $\bm{U}^*\in\Sigma_1$, its Taylor series can be written as
\begin{equation}
\mathcal{F}(\bm{U}) = \mathcal{F}(\bm{U}^*) + \bm{n\cdot \delta U} + \mathcal{O}(\bm{\delta U}^2)
\label{eq-taylor}
\end{equation}
with the displacement $\bm{\delta U}=\bm{U}-\bm{U}^*$ and the vector $\bm{n}$ with the components
\begin{equation}
n_i =  \frac{\partial \mathcal{F}}{\partial Y_a} \bigg|_{\bm{Y}=\bm{U}^*}~.
\end{equation}
For small displacements $\|\bm{\delta U}\|\ll 1$ higher order terms in Eq.~\eqref{eq-taylor} can be ignored.
From construction we know that $\mathcal{F}(\bm{U}^*)\approx 1$ for any $\bm{U}^*\in\Sigma_1$, since $f=1$.
Note that the vector $\bm{n}$ is the normal vector of the first unitary limit surface at the point $\bm{U}^*$.
This is because $\mathcal{F}(\bm{U})$ is invariant under infinitesimal, orthogonal displacements
$\bm{\delta U}\bot\ \bm{n}$, {\em i.e.} tangential displacements to $\Sigma_1$.

\begin{figure}[t]
\includegraphics[width=\columnwidth]{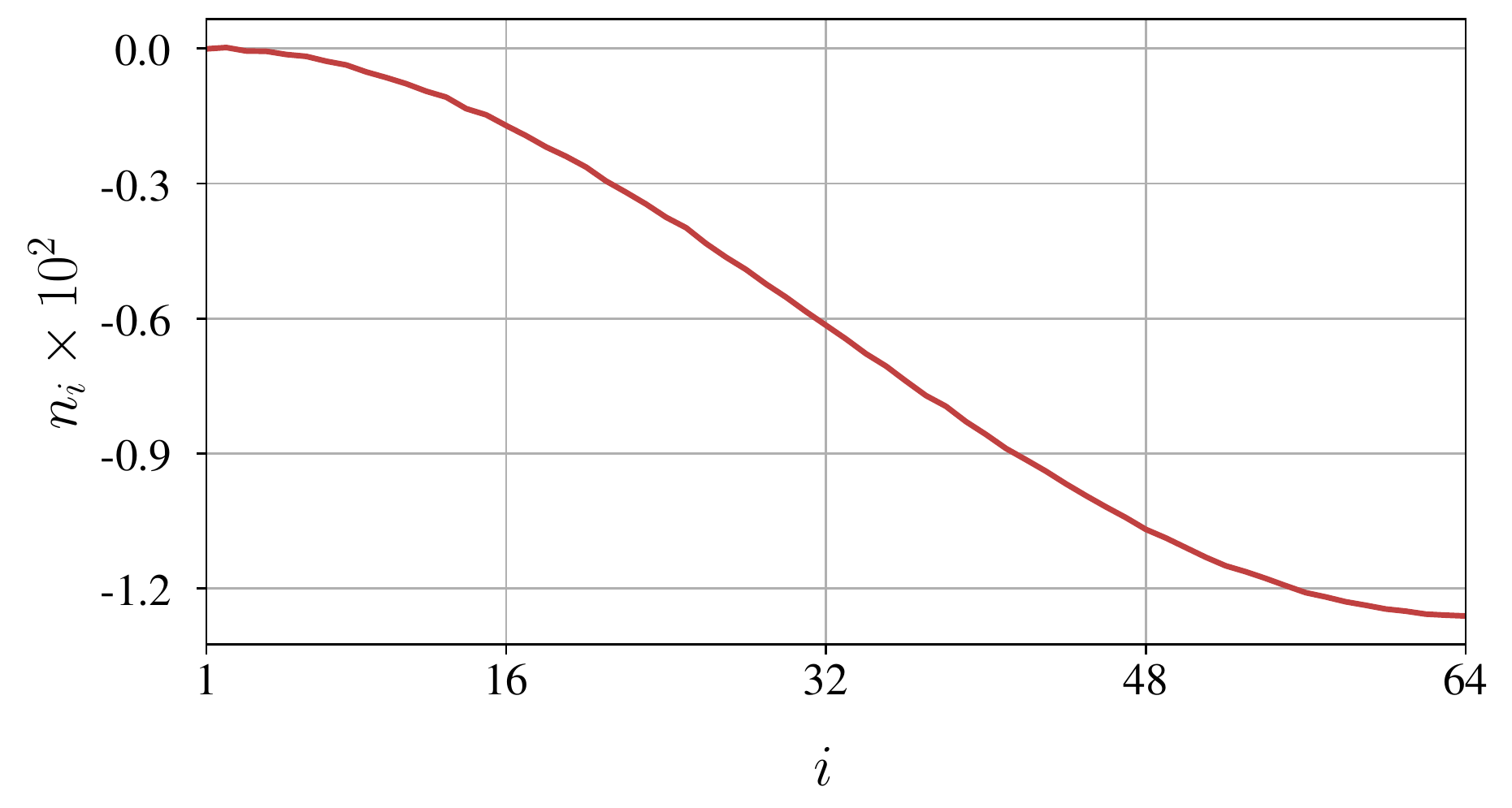}
\caption{Components of the normal vector $\bm{n}$ of the unitary limit surface at $\bm{U}^*=\pi^2/4\begin{pmatrix}
1,\ldots,1\end{pmatrix}$. As a gradient of $\mathcal{F}$, this vector points towards the strongest ascent of $f$,
which explains why its components are not positive.}
\vspace{-2mm}
\label{figSM-4}
\end{figure}\noindent
To give an example, we consider the first order Taylor approximation for the expansion point
$\bm{U}^*=\pi^2/4\begin{pmatrix} 1,\ldots,1 \end{pmatrix}$. This corresponds to an expansion around the
potential well with exactly one bound state that is also a zero-energy bound state. At first we derive
$\mathcal{F}(\bm{U}^*)=1-3.2\times 10^{-5}\approx 1$. Due to the rather involved architecture of
$\mathcal{F}$, we decide to calculate derivatives numerically,
\begin{equation}
\vspace{-2mm}  
n_i \approx \frac{\mathcal{F}(\bm{U}^*+\Delta \bm{e}_i)-\mathcal{F}(\bm{U}^*)}{\Delta}
\vspace{-1mm}
\end{equation}
with the $i^\text{th}$ basis vector $\bm{e}_i$ and the step size $\Delta=0.01$. The resulting components of the
normal vector are depicted in Fig.~\ref{figSM-4}. In this case, we can clearly see that $\bm{n}$ is far
from collinear 
to $\bm{U}^*$, which indicates that $\Sigma_1$ may have a rather complicated topology. Using the expansion in 
Eq.~\eqref{eq-taylor} and the components $n_i$ shown in Fig.~\ref{fig-4}, we arrive at an interesting and 
interpretable approximation of $\mathcal{A}(\bm{U})$ around $\bm{U}^*$,
\begin{equation}
\mathcal{A}(\bm{U})\approx -\frac{\mathcal{B}(\bm{U}^*)}{\bm{n\cdot \delta U}}~, \label{eq-aapprox-1}
\end{equation}
which lets us model the unitary limit in terms of a scalar product. Let us consider displacements 
$\bm{\delta U}=$~$(u$$-$$\pi^2/4)$$\begin{pmatrix}1,\ldots,1\end{pmatrix}$ that are parallel to $\bm{U}^*$. 
Inserting the value $\mathcal{B}(\bm{U}^*) = 0.81$ and $\sum_in_i=-0.40$, 
Eq.~\eqref{eq-aapprox-1} becomes
\begin{equation}
\mathcal{A}(\bm{U})\approx \frac{2.01}{u-{\pi^2}/{4}}~. \label{eq-aapprox-2}
\end{equation}
We can compare this to the expected behavior $a_0$$=$$1$$-$$\tan\sqrt{u}/\sqrt{u}$ of the S-wave scattering length 
for potential wells. The Pad\'e-approximant of order $[0/1]$ of this function at the point $u=\pi^2/4$ is given by
\begin{equation}
a_0\approx \frac{2}{u-{\pi^2}/{4}}~,
\end{equation}
which agrees very well with the approximation Eq.~\eqref{eq-aapprox-2} of scattering length predictions for inputs
$\bm{U}=~u\begin{pmatrix}1,\ldots,1\end{pmatrix}$ in the vicinity of $\bm{U}^*$ by the
quotient $\mathcal{A}$.

\end{document}